\documentclass[12pt]{article}

\begin{document}
\begin{flushright}
\end{flushright}

\begin{center}
\begin{large}
\textbf{
Towards Resolution of Hierarchy Problems
in a Cosmological Context
}

\end{large}

\vspace{2cm}
\begin{large}
M. Yoshimura

Department of Physics, Okayama University \\
Tsushima-naka 3-1-1 Okayama
Japan 700-8530
\end{large}
\end{center}

\vspace{4cm}

\begin{center}
\begin{Large}
{\bf ABSTRACT}
\end{Large}
\end{center}

A cosmological scenario is proposed,
which simultaneously solves the mass hierarchy and the small
dark energy problem.
In the present scenario an effective gravity mass scale
(inverse of the Newton's constant) increases during the inflationary period. 
The small cosmological constant or the dark energy density in the present
universe is dynamically
realized by introducing two, approximately O(2) symmetric dilatons,
taking the fundamental mass scale at $TeV$. 

\newpage
{\bf 1.
Introduction}

There are many interesting ideas that attempt to solve
the hierarchy
problem between gravty and particle physics mass scales
(the first hierarchy problem),
but none seems to have linked this hierarchy problem
with another hierarchy in cosmology (the second hierarchy problem);
presence of a finite, but very small cosmological constant, or
a dark energy, its nature and origin yet to be identified.
We attempt to construct models that simultaneously solve these hierarchy
problems by radically changing
cosmology in the same spirit of ideas as due to Dirac
\cite{dirac}, Brans and Dicke \cite{brans-dicke}.
Important new ingredient in the present work
is a choice of the dilaton potential
along with a curvature coupling similar to the one given by Brans and Dicke.

Recent observations of WMAP and the large scale structure confirm
the basic validity of the inflationary paradigm \cite{inflation}, 
but at the same time
it has left behind a great conundrum of the presence of the dark energy
which is close to, but dominant over, the dark matter energy.
The implied mass scale of (dark energy density)$^{1/4}$ is very small of
order $10^{-3} eV$ in the microscopic scale.
It thus appears that a resolution of great mysteries in cosmology via
inflationary scenarios 
has created another great mystery, which seems even more insurmountable.

We propose a possible scenario towards resolution of two
hierarchy problems, 
while retaining nice features of the inflationary cosmology.
Inflation is achieved by the dilatonic inflaton in our scenario.
We avoid fine tuning of parameters, taking a common mass scale of order $TeV$
for the dilatonic inflaton potential.
In a broken symmetric model of two dilatons a light scalar boson
of mass $\approx TeV^2/m_{pl} \approx 1 meV$ is predicted,
whose coupling to matter is gravitationally suppressed.

\vspace{1cm}

{\bf 2.
Theoretical framework}

We work in a general framework of four dimensional Lagrangian field theory,
with two parts left unspecified for the time being;
\begin{equation}
{\cal L} = \sqrt{-g}\,[-f(\varphi_i) R + \frac{1}{2}(\partial \varphi_i)^2
- V(\varphi_i) + {\cal L}_{m} ] \,.
\end{equation}
The dilatonic
coupling of the scalar field $\varphi_i$ to the scalar curvature given by
$f(\varphi_i) R$ is
taken from \cite{brans-dicke}.
But
we depart in the choice of the potential $V(\varphi_i)$ 
from the Brans-Dicke theory,
in which a single dilaton was introduced along with  the null potential
and $f(\varphi) = \epsilon \varphi^2$.

The Einstein gravity equation is modified to \cite{zee}
\begin{equation}
R_{\mu \nu} - \frac{1}{2}g_{\mu \nu}R = 
\frac{1}{2f}[T^{(m)}_{\mu \nu} + T^{(\varphi)}
_{\mu \nu}] + \frac{1}{f}
(f_{;\mu ;\nu} - g_{\mu \nu} f^{; \lambda}_{; \lambda}) 
\,,
\label{modified einstein}
\end{equation}
where $T^{(\varphi)}_{\mu \nu}$ is contribution to
the energy-momentum tensor from the scalar $\varphi$, while
$T^{(m)}_{\mu \nu}$ is the usual contribution of radiation, matter
and other fields.
Scalar field evolution is given by
\begin{equation}
\varphi^{; \lambda}_{i ; \lambda} = 
-\frac{\partial V}{\partial \varphi_i} - \frac{\partial f}{\partial \varphi_i}
R
\,.
\label{scalar-eq}
\end{equation}
One may use 
\begin{equation}
- R = \frac{1}{2f(\varphi)}[T - (\partial \varphi_i)^2 + 4V(\varphi_i)
- 6 f^{;\rho}_{;\rho}] \;, \;
\end{equation}
in the right hand side of eq.(\ref{scalar-eq}).
Here $T$ is the trace of the matter energy-momentum tensor.

An effective gravitational strength is given by
$f(\varphi_i) = 1/16\pi G$, and this can be spacetime dependent
due to a nontrivial spacetime dependence of $\varphi$, thus
modifying the Einstein equation in an essential way.
The usual Einstein equation with a constant $\varphi_i$
is however an excellent approximation in the present universe.
(We shall discuss its possible variation at the end of
this paper.)
Existence of the term, 
$(f_{;\mu ;\nu} - g_{\mu \nu} f^{; \lambda}_{; \lambda})/f$, 
in the modified Einstein equation (\ref{modified einstein})
is important in our cosmological discussion.
From reasons to be clarified later, 
we introduce two (or more) dilatons and extend the dilatonic coupling to
\begin{equation}
f(\varphi_i) = \epsilon_1 \varphi_1^2 + \epsilon_2 \varphi_2^2 \,,
\label{dilaton coupling}
\end{equation} 
with $\epsilon_i$ positive numbers. 
We later mention what happens in the case of the exact
$O(2)$ symmetry of $\epsilon_1 = \epsilon_2$.

We do not assume any fine tuning of the potential $V(\varphi_i)$ 
except that it is a bounded function allowing infinitely many negative values
and  infinitely many local minima. 
In this way a large mass hierarchy and 
dynamical relaxation towards a small cosmological constant may be realized.
The simplest choice realizing these is a periodic
potential of minimum numbers of parameters;
\begin{equation}
V(\varphi_i) = V_{0}\cos \frac{\varphi_r}{M} + \Lambda
\,,
\label{sinusoical}
\end{equation}
with $\varphi_r = \sqrt{\varphi_i^2}$ and $V_0 > \Lambda > 0$.
Here $\Lambda$ is a collection of all constants in the standard model
Lagrangian ${\cal L}_m$ such that the potential of the standard
model Lagrangian vanishes at its minimum.
We assume O(2) rotational symmetry for the potential $V(\varphi_i)$.

Important features of our assertions below are valid irrespective of the
precise form of the potential. 
The essential requirement on the potential for a successful scenario is
that (1) boundendness, (2) infinitely many local
minima, and (3) infinitely many regions of negative values 
between minima and maxima.
Nevertheless, it would be useful to have
a simple realization such as (\ref{sinusoical}) 
of our idea and to discuss a model explicitly.

For both simplicity and naturalness we assume that all mass parameters
are of the same order, thus
$V_{0} \approx \Lambda = O[M^4]$ for the choice (\ref{sinusoical}).
We take the common mass scale $M$ of order $TeV$.

\vspace{1cm}
{\bf 3. Dynamical equation}

Consider the Robertson-Walker metric of flat universe,
\( \:
ds^2 = dt^2 - a(t)^2 d \vec{x}^2 \,.
\: \)
Dynamical equation of time evolution is derived straightforwardly.
We write it down in terms of the following two field variables
$f_{\pm}$,
\begin{equation}
\varphi_1 = \sqrt{\frac{f_+}{2\epsilon_1}}
\,, \hspace{0.5cm}
\varphi_2 = \sqrt{\frac{f_-}{2\epsilon_2}}
\,, \hspace{0.5cm}
f_{\pm} = f \pm k
\,.
\end{equation}
The basic dilaton dynamics is given by
\begin{eqnarray}
&&
\hspace*{-1cm}
\ddot{f_+} + 3H \dot{f_+}  = 
[1 + 6\frac{\epsilon_1 f_+ + \epsilon_2 f_-}{f}]^{-1}
[4\epsilon_1 f_+ (- \tilde{R} - \frac{1}{2\epsilon_1 \varphi_r}V')
+  \frac{\dot{f_+}^2}{2f_+}]
\,,
\\ &&
\hspace*{-1cm}
\ddot{f_-} + 3H \dot{f_-}   = 
[1 + 6\frac{\epsilon_1 f_+ + \epsilon_2 f_-}{f}]^{-1}
[4\epsilon_2 f_- 
(- \tilde{R} - \frac{1}{2\epsilon_2 \varphi_r}V')
+ \frac{\dot{f_-}^2}{2f_-}]
\,,
\\ &&
\tilde{R} = \frac{1}{2f}(\frac{\dot{f_+}^2}{8\epsilon_1 f_+} 
+ \frac{\dot{f_-}^2}{8\epsilon_2 f_-} - 4V - T) 
\,.
\label{dilaton correction}
\end{eqnarray}
These ought to be solved along with the modified Einstein equation,
\begin{eqnarray}
&&
H^2 =
\frac{1}{6f}(T_{00} + \frac{1}{2}\dot{\varphi_i}^2
+ V) - H\frac{\dot{f}}{f}
\,.
\label{hamiltonian constraint}
\end{eqnarray}

We note interesting features of $\varphi$ dynamics.
First, the usual force term 
$-\,\frac{\partial V}{\partial \varphi}$ is modified,
as seen from the $f$ dependent terms of (\ref{dilaton correction}) 
$- 2f_{\pm}V/f$
present in $f_{\pm}$ equations.
The second is presence of the induced matter coupling
$\propto T$, which may
be derived from an effective Lagrangian of the form,
$\frac{T}{2} \ln f$.
Thus, the dilaton couples to matter via
the trace of the energy-momentum tensor $T$.

One dilaton model fails to solve hierarchy problems from a number of reasons;
it requires a fine tuning at the stationary minimum of
the potential, namely $V=0$ at the same time when
$- V_{,\, \varphi} + 4V/\varphi = 0$.
It is also difficult to obtain a small dark energy density.

The key for success is introduction of more freedom such as an angular
momentum in a higher dimensional $\varphi$-space.
We thus introduce 2 dilatonic inflatons $\varphi_{1} \,, \varphi_{2}$, 
as already stated.
This makes it possible to reach a large $f$ value (very weak
gravitational interaction) without being trapped in many
potential minima of negative cosmological constant.

\vspace{1cm}
{\bf 4.
Cosmological evolution
}

Let us first point out that this model realizes the power-law
inflation.
Ignoring, for the moment, potential variation and replacing $V$ 
by its averaged value 
\(\:
\Lambda ,
\:\)
we seek solution, with the ansatz valid for large $t$,
\begin{equation}
f = At^2
\,, \hspace{1cm}
k = Bt^2
\,, \hspace{1cm}
a \propto t^{\omega}
\,.
\end{equation}
Leading order solution is found, in which $\omega$
is determined in terms of $\epsilon_i$. In the small $\epsilon_i$
limit the index of the power $\omega$ becomes large.

A large value of $\omega$ is favored to approximately mimic
the exponential expansion of the cosmological scale factor.
The gravity mass scale increases as inflation proceeds like
\(\:
f \propto a^{2/\omega} \,.
\:\)
Thus, there is no difficulty of obtaining a large enough e-folding
factor of inflation, at the same time resolving the mass hierarchy
problem.
This is a feature already visible in the model of extended inflation
\cite{extended inflation}, although it has not been much appreciated.

After this inflationary epoch, the inflaton $\varphi_i$ is
expected to settle down to some stationary points.
But in our model of $\epsilon_1 \neq \epsilon_2$ 
there is no stationary point,
because the requirement of constant $\varphi_i$ values implies
both $V' =0$ and $V=0$, which is nothing but the fine tuning of parameters of
the potential.
The model without fine tuning however gives a mechanism of
dynamical cancellatin of the effective cosmological constant,
\(\:
\Lambda_{eff} =
\langle \frac{1}{2}\dot{\varphi_i}^2 + V
\rangle .
\:\)
We shall discuss this mechanism later.

Particle production right after inflation gives rise to
the hot big bang. We shall briefly discuss how this comes about.
We take an example of the matter energy-momentum tensor,
\(\:
T = \frac{1}{2}m_{\psi}^2 \psi^2
\,,
\:\)
where $\psi$ represents a generic boson.
The mode equation for $\psi$ when $f$ deviates around $1/16\pi G$
by a small amount is
\begin{eqnarray}
&&
\ddot{\psi} + (m_{\psi}^2 + k^2) \psi + 3H\dot{\psi}
+ \tilde{m}_{2}^2 \xi(t) \psi = 0
\,,
\\ &&
\xi(t) =
\frac{m_{\psi}^2}{2\tilde{m}_{2}^2}(\frac{\delta f}{f} - 
\frac{1}{2}(\frac{\delta f}{f})^2 + O[(\frac{\delta f}{f})^3])
\,.
\end{eqnarray}
There exist two mass eigenstates of the dilaton, and we took
the lighter dilaton since it gives more important contribution here.

We shall first discuss the fluctuation equation in order to derive
these mass values.
With the ansatz,
\(\:
f = f_0 + \delta f
\,, \hspace{0.3cm} 
k = k_0 + \delta k
\:\)
the linearized equations follow
\begin{eqnarray}
&&
(
\begin{array}{c}
  \ddot{\delta f_+ } + 3H \dot{\delta f_+} \\
  \ddot{\delta f_- } + 3H \dot{\delta f_-}
\end{array}
)
=
{\cal M}^2
(
\begin{array}{c}
 \delta f_+ \\  \delta f_-
\end{array}
)
\,,\hspace{1cm}
{\cal M}^2_{ij} = O[M^2]
\,, 
\\ &&
det ({\cal M}^2)
\approx
\frac{V' V''}{\varphi_r^3 }(\frac{f_+ }{\epsilon_1} + 
\frac{f_- }{\epsilon_2})
\,,
\end{eqnarray}
where $det ({\cal M}^2)$ appears $O[M^5 m_{pl}^{-1}]$, 
but actually $O[M^6 m_{pl}^{-2}]$,
since $V' \propto 1/m_{pl}$, as is shown later.

The mass diagonalization, taking into accout $V' \propto 1/m_{pl}$, 
yields two eigenmasses of order,
\begin{equation}
\tilde{m}_1 = O[M]
\,, \hspace{0.5cm}
\tilde{m}_2 = O[\frac{M^2}{m_{pl}}] = O[1 meV] (\frac{M}{ 1 TeV})^{2}
\,.
\end{equation}
We call the second, light dilaton newdiron, named after Newton and Dirac.

Back to the inflaton decay,
large and small amplitude decay of $\varphi$ is described as follows.
First, the large initial amplitude condition is fullfilled,
since
\begin{equation}
\xi(t_i) = O[\frac{m_{\psi}^2 M }{\tilde{m_1}^2 m_{pl}}] =
O[\frac{m_{\psi}^2 m_{pl}}
{M^3 } ] \gg 1
\,,
\end{equation}
for $\tilde{\delta f_2}$ decay, taking initially
\(\:
\delta \varphi_i (t_i) = O[M] .
\:\)
Thus, the inflaton oscillation leads to explosive particle production due to
the parametric resonance effect \cite{preheating}.
Thermalization is quickly achieved, giving a reheat temperature of order
$M$.

After this initial phase of preheating and
thermalization, the dimensionless amplitude $\xi(t)$ 
drops to $O[1]$, and the large amplitude oscillation
stops. After this takes place, the only process of dilaton
decay is two-body perturbative decay.
To correctly derive these rates, one has to note correctly
normalized fields given by
\(\:
(2\sqrt{2\epsilon_i})^{-1}\delta f_{\pm}/\sqrt{f_{\pm}} \,.
\:\) 
These decay rates are then
\(\:
\Gamma_1 = O[M^3 /m_{pl}^2]
\,,
\:\)
\begin{equation}
\Gamma_2 = O[\frac{m_{\psi}^4}{m_{pl}^2 \tilde{m}_2}] = 
O[\frac{M^6}{m_{pl}^5}] \approx 
(10^{48} sec)^{-1}(\frac{M}{1 TeV})^{6}
\,,
\end{equation}
where for the newdiron only neutrino-pair and two photon
decays are possible due to the small mass.
Thus, the newdiron is effectively stable.

For the success of nucleosynthesis,
the heavy dilaton must decay prior to nucleosynthesis.
This gives a mass scale constraint;
\begin{equation}
\Gamma_1 = O[(10^3 sec)^{-1}(\frac{M}{1 TeV})^3] 
> (10^3 sec)^{-1}
\; \Rightarrow \;
M > {\rm a \; few\;} TeV
\,.
\end{equation}

\vspace{1cm}
{\bf 5. Dynamical relaxation towards vanishing $\Lambda_{eff}$}

At some stage of cosmological evolution after inflation,
the cosmological constant (or more precisely the vacuum energy density)
of order $(TeV)^4$
must relax towards smaller values of order $(meV)^4$.
We analyze this problem by assuming the main terms 
$f$ and $ k$ of order $ f_0 $ (constant) and
$\dot{\delta k} \gg \dot{\delta f}$.
The resulting equation for $k$ fluctuation is
\begin{eqnarray}
\ddot{\delta k} + 3H \dot{\delta k} 
= (1 + 6\frac{\epsilon_1 f_+ + \epsilon_2 f_- }{f_0})^{-1}
[- \frac{2V'}{\varphi_r}(k_0 + \delta k) + \frac{4}{f_0}(\epsilon_1 f_+ 
- \epsilon_2 f_-)V]
\,.
\nonumber 
\end{eqnarray}
From the solution of
\(\:
\delta k = O[m_{pl}] \times
\:\)
oscillating function, one obtains the kinetic term of order,
\begin{eqnarray}
\frac{\dot{\varphi_k}^2}{2}
= O[1] \times
(\int^t dt' V'(t') )^2 = O[M^4]
\,.
\label{kinetic}
\end{eqnarray}
Since this quantity is nonvanishing and of order $M^4$, there
is a chance of cancellation against the potential term $V$.
Indeed, for a large $f_0$ there are a great many trajectories of
vanishing $\Lambda_{eff}$.

The requirement of constant $f$ (gravity scale) and 
the tuning condition $\Lambda_{eff} = 0$  give
\begin{equation}
\frac{\dot{\varphi_k}^2}{2} + V = 0
\,, \hspace{0.5cm}
3V - 
\frac{\epsilon_1 + \epsilon_2}{4\epsilon_1 \epsilon_2}\frac{f_0}{\varphi_r}V' 
= 0
\,.
\label{fine tuning}
\end{equation}
These two equations (\ref{fine tuning}), 
to be supplemented by eq.(\ref{kinetic}),
ought to be solved in favor of $(f_0,\, k_0)$.
Suppose that the Newton's constant is measured to a precision $\delta$
(which is at present $\approx 10^{-4}$). Then, there are of order
$10^{15} \delta (M/1 TeV)^{-1}$ possible $(f_0,\, k_0)$ values that
fit measured data.

Note that the coefficient of these eq.(\ref{fine tuning}),
$f_0/\varphi_r$ is of order $m_{pl}$.
Hence the factor $V'$
must be small of order $M^4/m_{pl}$ to cancel against other terms of order
$M^4$.
This explains the already mentioned mass relation,
$\tilde{m_2} \approx M^2/m_{pl} $ due to the presence of 
$V'$.

There are a great many $(f,\,k) $ values of solutions to both of
eqs.(\ref{fine tuning}), close to trajectories of
$V' = 0$, or more precisely
\( \:
\varphi_r = M N\pi + O[M^2/m_{pl}] 
\:\)
for a large integer $N$.
Thus, a very small $\Lambda_{eff}$ may be dynamically obtained near these
points which however cannot be stationary anchor points, hence one
expects that never-ending shifts towards these points occur
in the present version of model of $\epsilon_1 \neq \epsilon_2$.

We however point out with the exact $O(2)$ symmetry of 
$\epsilon_1 = \epsilon_2 = \epsilon$ there exist stationary anchor
points of very small effective cosmological constant.
This comes about, because the angular momentum $L = \varphi_1 \dot{\varphi_2}
- \varphi_2 \dot{\varphi_1}$ is conserved with the $O(2)$ symmetry, and
the stationary condition reduces to
\begin{equation}
- V' + \frac{4V}{\varphi_r} + \frac{L^2}{8\epsilon \varphi_r^3} = 0
\,, \hspace{1cm}
\frac{L^2}{2 \varphi_r^2} + V = 0
\,,
\end{equation}
which should be solved for $(f_0,\, L)$.
There are again such candidate values of order $10^{15} \delta (M/1 TeV)^{-1}$ 
that fit observation.
The critical question for realization of this result concerns
a natural initial setting for the conserved quantity 
$L^2$ which should be of order $M^4 m_{pl}^2$.
We shall address this question elsewhere.

The $O(2)$ symmetric model has one heavy dilaton of mass $ O[M]$,
whose decay rate is of order $M^3 /m_{pl}^2$.
What happens to the light dilaton is that it becomes massless, which however
completely decouples from the rest of the world.

In both scenarios of dynamical relaxation towards the vanishing 
cosmological constant, the exact tuning is not necessary and 
moreover is unlikely to occur. 
Under this circumstance
one expects a residual dilaton energy as the dark matter candidate.
The tuning to the amplitude precision of the leading order $M^2/m_{pl}$ yields
the dark energy density of order,
\begin{equation}
\rho_{DM} = O[(\frac{M^2}{m_{pl}})^4] (\frac{T_0}{T_d})^3 
\,,
\end{equation}
with $T_d /T_0$ the expansion factor after the relaxation.
If the relaxation epoch is close to the present age, the dilaton
oscillation energy is of the right order of magnitude to
explain the present amount of dark matter.

There is another possibility.
Suppose that the approach to anchor points had occurred around nucleosynthesis
or at the heavy dilaton decay. Then, one gets a right order of magnitude of
the present $\Lambda_{eff}$, since its present value
\begin{equation}
(\delta \Lambda_{eff})_0 = O[\frac{M^6}{m_{pl}^2}] (\frac{T_0}{MeV})^3
\approx (1 meV)^4 (\frac{M}{1 TeV})^6
\,.
\end{equation}

Finally, let us discuss what happens if the scenario works,
as expected.
The result differs, depending whether the dark matter is provided
by the dilaton oscillation or another form of stable particles such
as lightest supersymmetric particles.
If the dark matter is made of newdiron, one obtains, from the
consistency with the modified Einstein equation, for the
$w$ value defined by
\(\:
w \equiv p/\rho ,
\:\)
\begin{equation}
w = - 1 - \frac{\rho_{DM}}{\rho_{DE}}
\,.
\end{equation}
If the dark matter is attributed to another source, one has
the usual $w = -1$.
Clearly, a better understanding of the relaxation process is welcome.
The time dependence of $w$ differs, depending on how the relaxation occurs,
in particular, when this occurs.
Observation of future deep sky surveys is crucial to test the model
of dynamical relaxation.

\vspace{1cm}
{\bf 6. Variation of Newton's constant}

In the present model variation of the gravitational constant is
inevitable, although its magnitude is model dependent.
We shall discuss the simplest case of how much it varies
due to nonrelativistic (NR) matter of mass density
$\rho_m$ excluding the dilaton oscillation.
For simplicity, we take the $O(2)$ symmetric model.
The quantity $f$ varies, with $m_d$ the heavy dilaton mass, according to
\begin{equation}
\ddot{\delta f} + 3H\dot{\delta f} + m_d^2 \delta f =
 \frac{2\epsilon}{1 + 12\epsilon}
\rho_m
\,.
\end{equation}
Asuuming that this NR matter dominates as the main component of
the dark matter, one derives a $\delta f/f$, hence $-\delta G /G$, 
variation between epochs of NR matter appearance and
NR matter dominance
\(\:
 \approx 16\epsilon/ ((1 + 12\epsilon)N_d) \,, 
\:\)
where $N_d$ is the relativistic degrees of freedom contributing to
the energy density at NR matter appearance.
Thus, this fraction can be made small, and furthermore its change after
NR matter dominance is small, although the variation can be made larger
to accommodate some nonstandard varying $G$.
We shall discuss elsewhere how much $G$ varies when the dark matter
is made of the newdiron.

\vspace{1cm}
In summary, the Dirac's large number hypothesis
has been resurrected along with inflation.
Furthermore, a class of multi-dilaton models
give a possibility of solving the problem of 
how the present dark energy density becomes of order $(TeV^2/m_{pl})^4$,

Interesting details and some extentions of the present model
will be presented in separate publication.

\end{document}